\documentclass[11pt]{iopart}
\usepackage{iopams}
\usepackage{txfonts}
\usepackage{graphicx}
\def\hi{H\,{\sc i}~}
\begin{document}

\title{Galactic outflows and the chemical evolution of dwarf galaxies}

\author{S. Recchi$^1$, G. Hensler$^1$, D. Anelli$^2$}

\address{$^1$ Institute of Astronomy, Vienna University, 
T\"urkenschanzstrasse 17, A-1180 Vienna, Austria}
\address{$^2$ Department of Physics, Trieste University, 
Via A. Valerio 2, 34127, Trieste, Italy}
\ead{simone.recchi@univie.ac.at}
\begin{abstract}

Galactic winds in dwarf galaxies are driven by the energy released by
supernova explosions and stellar winds following an intense episode of
star formation, which create an over-pressured cavity of hot gas.
Although the luminosity of the star formation episode and the mass of
the galaxy play a key role in determining the occurrence of the
galactic winds and the fate of the freshly produced metals, other
parameters play an equally important role.  In this contribution we
address the following questions $(i)$ What is the late evolution of
superbubbles and what is the final fate of the superbubble cavities?
$(ii)$ How does the multi-phase nature of the ISM, in particular the
coexistence of hot gas with embedded clouds, affect the development of
galactic winds?  $(iii)$ What is the relation between the flattening
of a galaxy and the development of bipolar galactic winds?

\end{abstract}

%\maketitle

\section{Introduction}

After an intense episode of star formation (SF), the interstellar
medium (ISM) in galaxies can be swept up by the mechanical energy of
multiple supernova (SN) explosions and stellar winds.  If the velocity
achieved by this swept-up ISM reaches the escape velocity of the
galaxy, we are in the presence of a {\it galactic wind}.  Large-scale
galactic outflows have been observed in the most active starburst
galaxies both in the local Universe (Dahlem et al. 1998) and at high
redshifts (Pettini et al. 2000).  Although the first theoretical
studies of galactic winds were applied to elliptical galaxies
(e.g. Mathews \& Baker 1971), it has long been argued that they are
more effective in dwarf galaxies (DGs) since their potential wells are
shallower and it is easier to overcome the escape velocity (Larson
1976).  It has also been speculated that this physical process
determines the mass-metallicity relation (Dekel \& Silk 1986).

If the ISM in a galaxy is stratified, the swept-up gas expands
preferentially along the steepest density slope, i.e. the direction
perpendicular to the galaxy plane.  A number of hydrodynamical
simulations of the development of galactic winds in stratified media
have been performed in the last two decades (i.e. MacLow \& Ferrara
1999, hereafter MF99; D'Ercole \& Brighenti 1999; Fragile et al. 2004
and references therein).  As a result of these simulations, some
consensus has been reached that galactic winds are not capable of
removing a large fraction of the ISM from a galaxy but they are able
to eject a significant fraction of the metal-enriched gas produced
during the SF episode (but see Silich \& Tenorio-Tagle 1998 for a
different view).  In particular, after the work of MF99 it is believed
that the DGs with masses smaller than $\sim$ 10$^8$ M$_\odot$ are able
to expel the majority of the metals freshly produced after the episode
of SF whereas only for galaxies as small as $\sim$ 10$^6$ M$_\odot$
the ejection of a large fraction of the ISM initially present inside
the galaxy is possible.

In spite of the important theoretical progress made on this field in
the last years, the theory of the galactic winds has proven to be much
more complex than depicted above and the results of MF99 must be seen,
in most of the cases, as oversimplified.  In this contribution we
analyze the following questions, not addressed in the previously
mentioned literature:

\begin{itemize}

\item What is the late evolution of superbubbles and what is the final 
fate of the superbubble cavities?

\item How does the multi-phase ISM, in particular the coexistence of hot 
gas with embedded clouds, affect the development of galactic winds?

\item What is the relation between the geometry of a galaxy (in 
particular its flattening) and the development of bipolar galactic
winds?

\end{itemize}

\section{The refill of superbubble cavities}

\begin{figure}[ht]
 \begin{center}
\hspace{1.5cm}\includegraphics[width=12cm]{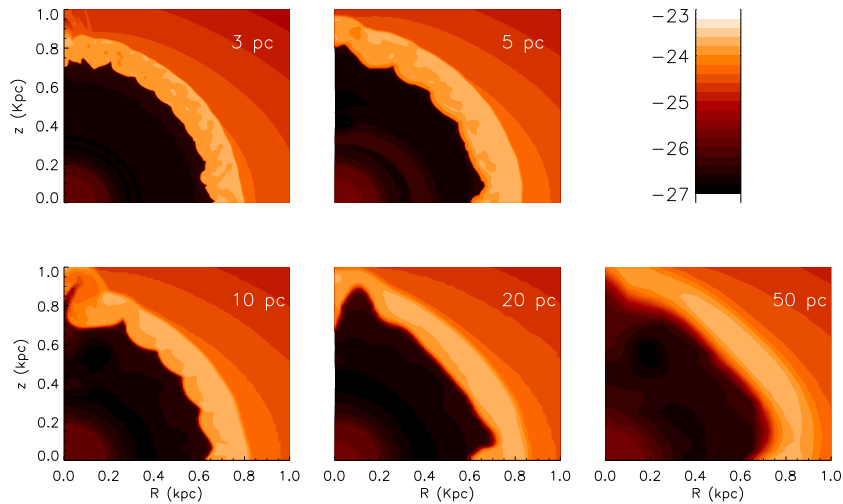}
\vspace{-1cm}
\caption{ Density contours after 50 Myr of evolution for model galaxies at 
different spatial resolutions (labeled in pc at the top right corner
of each panel).  The density scale (in g cm$^{-3}$) is on the upper
right panel.}
\label{resol} % for cross-references 
\end{center}
\end{figure}

After the energetic SN explosions and stellar winds have swept up the
ISM of a galaxy, a large cavity filled with hot gas is left behind.
The energy input rate after an episode of SF declines with time,
therefore, the hot cavity loses pressure and the supershell tends to
recede towards the center of the SF region.  Superbubbles solely
produced by SNeII experience a buoyancy of the hot gas at the end of
the SF phase and the central region can be completely refilled with
cold gas in a timescale of the order of $\sim$ 10$^8$ yr or less
(D'Ercole \& Brighenti 1999).  However, the occurrence of SNeIa, which
explode with much longer delays compared to the SNeII (see
e.g. Matteucci \& Recchi 2001) can change the thermodynamical behavior
of the gas in the center of galaxies.  By means of detailed numerical
simulations we have analyzed the late evolution of a galaxy after a
single SF episode in order to explore the timescale needed to refill
the hot cavity, taking into account both SNeII and SNeIa (Recchi \&
Hensler 2006).  Although our simulations are not adaptive, we have
performed a thorough spatial resolution study (see an example of
compared snapshots at different resolutions in Fig. \ref{resol}) and
we have verified that at a resolution of $\sim$ 10 pc the main
physical phenomena included in the code are numerically properly
treated.  This is therefore the average numerical resolution we have
used in our simulations.

\begin{figure}[ht]
 \vspace{-5cm}
 \begin{center}
\hspace{1.5cm}\includegraphics[width=12cm]{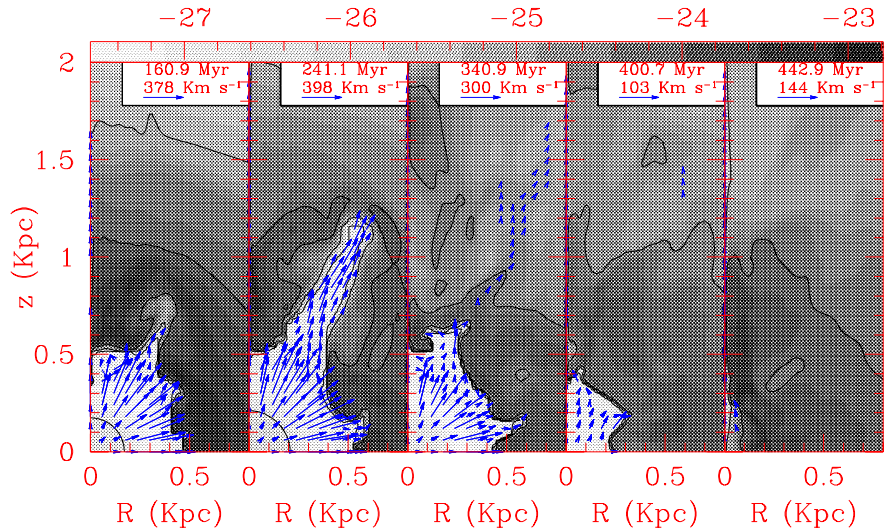}
\caption{ Density contours and velocity fields for a model illustrating 
the process of refilling the superbubble cavities (see text) at five
different epochs (evolutionary times are labelled in the box on the
upper right corner of each panel).  The density scale is given in the
strip on top of the figure.  }
\label{HL} % for cross-references 
\end{center}
\end{figure}

As a typical evolutionary sequence of the refill process we show in
Fig. \ref{HL} snapshots of a model with a moderate SF episode (0.05
M$_\odot$ yr$^{-1}$) lasting 200 Myr.  The combined energy of SNeIa
and SNeII is able to break-out of the galaxy at t $\sim$ 160 Myr
(first panel).  The superbubble starts funneling through the \hi due
to stratification of the ISM.  After the additional time interval
during which SNeII explode, i.e. after $\sim$ 230 Myr, SNeIa still
provide enough energy to sustain the outflow (panel 2).  The funnel
begins to shrink at t $\sim$ 300 Myr and at $\sim$ 340 Myr the outflow
has almost completely disappeared (panel 3).  At $\sim$ 400 Myr the
cavity has approximately the original size of the SF region (panel 4)
and from now on it recedes further towards the center, therefore the
refill timescale of this model is $\sim$ 200 Myr.  The refill of the
cavity is mostly due to the pressure gradient created after the
superbubble breaks out the disk.

The refill timescale however strongly depends on the amount of gas
initially present inside the galaxy and on the duration and intensity
of the SF episode and ranges between 125 (model with a large initial
amount of \hi and short and intense SF) and 600 Myr (model with a
smaller initial \hi mass and milder and long SF).  This means that
large \hi holes (like the ones observed in many dIrr galaxies) can
survive a few hundred Myr after the last OB stars have died.

A SF occurring in the refilled cavity would produce metals which mix
with the surrounding unpolluted medium in a timescale of the order of
10--15 Myr, therefore any further episode of SF would stem out of a
metal-enriched ISM.  This is at variance with what happens if the
center of the galaxy is still occupied by hot and diluted gas because,
in this case, most of the metals are either directly ejected outside
the galaxy through galactic winds or are confined in a too hot medium,
therefore cannot contribute to the chemical enrichment of the warm
ionized medium (see e.g. Recchi et al. 2006).  More details can be
found in Recchi \& Hensler (2006).

\section{The effect of clouds in a galactic wind on the evolution of 
gas-rich dwarf galaxies}

Although the multi-phase nature of the ISM, in particular its
clumpiness, is observationally well established in DGs (Cecil et
al. 2001; Leroy et al. 2006), most of the hydrodynamical simulations
of galactic winds have focused on flows in homogeneous media, although
several attempts to perform multi-phase hydrodynamical simulations
have been made in the past, particularly, using the so-called {\it
chemodynamical} approach (Theis et al.  1992; Hensler et al. 2004).

We simulated models with structural parameters similar to the
well-known gas-rich DGs I Zw 18 and NGC 1569.  These galaxies have
been already modelled (Recchi et al. 2004; 2006) but considering only
a diffuse medium, without clouds.  We increased arbitrarily the gas
density of some specific regions of the computational grid, in order
to create a ``cloudy'' phase.  We either perturbed the initial gaseous
distribution or continuously created clouds, at a rate which equals
the SF rate, giving them an infall velocity of 10 km s$^{-1}$ along
the polar direction.  We addressed the question how and to which
extent this ``cloudy gas phase'' alters the results obtained in the
above mentioned papers, in particular for what concerns the
development of galactic winds and the chemical evolution of the
galaxy.

\begin{figure}
 \begin{center}
 \hspace{1.5cm}\includegraphics[width=12cm]{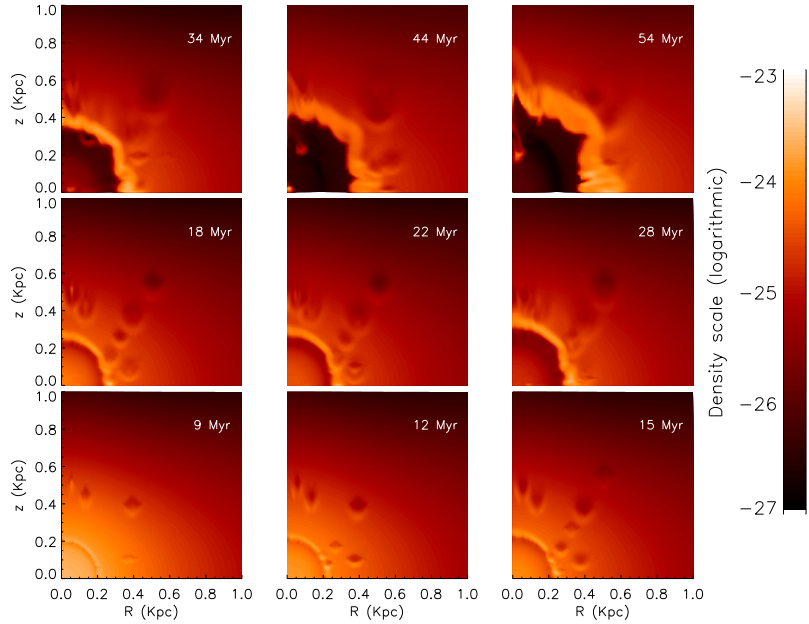}
\caption{ Density contours for a model of I Zw 18 with 
infalling clouds at 9 evolutionary times (labeled in Myr at the top
right corner of each panel). The density scale is on the right-hand
strip.  }
\label{9s} % for cross-references 
\end{center}
\end{figure}

The clouds are subject to a variety of disruptive phenomena like
evaporation, formation of shocks, development of dynamical
instabilities (in particular the Kelvin-Helmholtz instability) and
expansion due to the larger pressure compared to the surrounding
interstellar medium, therefore their average lifetime is relatively
short.  Nevertheless, they affect significantly the dynamical and
chemical evolution of a galaxy.  In fact, the clouds, when they
evaporate inside the superbubble, produce mass loading, increase the
mean density of the cavity gas and, therefore, enhance the radiative
energy losses.  This results in a significant decrease of the total
thermal energy (of the order of $\sim$ 20 -- 40\% compared to the
diffuse models), therefore less energy is available to drive the
development of a large-scale outflow.

On the other hand, the relative motion of supershell and clouds can
structure and pierce the expanding supershell, in particular, in
models with the setup of infalling clouds.  By this, holes and fingers
are created, as perceivable in Fig. \ref{9s}.  These holes destroy the
spherical symmetry initially present and favor the rushing out of the
highly pressurized gas contained in the cavity.  Therefore, in spite
of the reduced thermal energy budget, the creation of large-scale
outflows is not suppressed but, in most of the explored cases, only
slightly delayed.  A complete suppression of the development of the
galactic outflow happens only of we consider a very large cloud
falling in the galaxy along the polar direction (Recchi et al. 2006).
The pressure inside the cavity is reduced compared to diffuse models,
therefore in any case the total amount of ejected pristine ISM
(i.e. the low-metallicity gas not produced by the ongoing SF) is very
small.  However, the piercing of the supershell can lead to an
ejection efficiency of freshly produced metals as high as the one
attained by diffuse models, therefore, the diminished thermal energy
of these models does not imply that a larger fraction of metals is
retained inside the galactic regions.  On the other hand, since the
clouds are assumed to have primordial chemical composition, their
destruction and mixing with the surrounding medium reduces the total
chemical composition without altering the abundance ratios (see also
K\"oppen \& Hensler 2005).  This produces a final metallicity $\sim$
0.2 -- 0.4 dex smaller than the corresponding diffuse models.  For
details see Recchi \& Hensler (2007).

\section{The effect of geometry}

The development of bipolar galactic winds is favored by a flat galaxy
geometry since, in this case, there is a direction in which the work
required to extract gas out of the galaxy potential well is
particularly small, whereas in spherical galaxies the energy produced
by the SF must be high enough to expel the gas in all directions.
Numerical simulations of the evolution of such spherical galaxies have
shown that the ISM pressure can confine the supershell inside the
galaxy (see e.g. Marcolini et al. 2006, Recchi et al. 2007).  Of
course, the flatter the galaxy, the steeper the pressure gradient
along the polar direction, and therefore the easier it is to produce
bipolar galactic winds.  This issue has been already analyzed by means
of numerical simulations (see e.g. Strickland \& Stevens 2000).
However there is no systematic study in the literature about the
effect of geometry on the development of galactic winds, neither it
has been analyzed how does it affect the chemical evolution of
galaxies.

\begin{figure}
 \begin{center}
 \hspace{1.5cm}\includegraphics[width=12cm]{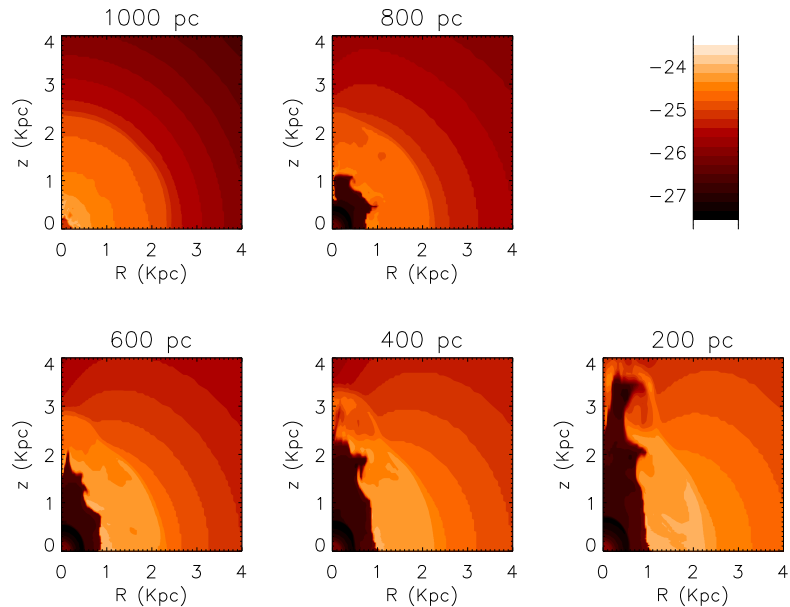}
\caption{ Density contours after 200 Myr of evolution 
for models differing on the semi-minor axis of their initial
configurations (semi-minor axis indicated on top of each panel). The
density scale is on the upper right panel.  }
\label{flat} % for cross-references 
\end{center}
\end{figure}

We have performed simulations of the evolution of DGs, with equal mass
of $\sim$ 4 $\cdot$ 10$^8$ M$_\odot$, whose initial gas distribution
is ellipsoidal, with constant semi-major axis (1 kpc) and variable
semi-minor axis.  In particular, we have analyzed 5 different values
of the semi-minor axis: 1 kpc (spherical model), 800 pc, 600 pc, 400
pc and 200 pc (flattest model).  For each model we have assumed a
constant SF rate of 0.025 M$_\odot$ yr$^{-1}$ and we have run the
simulation for 500 Myr.  In Fig. \ref{flat} we show the density
distribution of gas for each of these models after 200 Myr of
evolution.  As expected, the flatter the galaxy is, the more intense
and extended the galactic wind is.  In particular, the spherical model
does not show any sign of outflow whereas the flattest model shows a
galactic wind extending up to $\sim$ 4 kpc in the polar direction.
However, also in the flattest model, the total amount of ISM expelled
from the galaxy at the end of the simulation is only $\sim$ 15 \%.

However, as we have seen in the previous sections, a large fraction of
the freshly produced metals can be channelled along the galactic wind,
therefore the ellipticity of the galaxy can have important
consequences on the chemical evolution.  The spherical model does not
produce outflows, therefore it retains all the produced metals.  For
the flattened models instead the loss of metals is significant and it
increases proportionally to the ellipticity of the parent galaxy.  The
flattest galaxy retains at the end of the simulation only about one
third of the metals produced during the SF episode.  We have also
considered models with short and intense SF (with a rate of 0.5
M$_\odot$ yr$^{-1}$ lasting for 25 Myr) and the results are
qualitatively the same, namely the spherical model does not produce
outflows whereas the flattened models show an ejection efficiency of
metals increasing with the ellipticity, with the flattest model losing
$\sim$ 60\% of the metals produced by the burst of SF through the
galactic wind.

\ack
We thank the organizers for putting together a very enjoyable
workshop.  Part of the work was funded by the Alexander von Humboldt
Foundation and by the DFG under grant HE 1487/28-1.  The FWF is
acknowledged for support with travel money.

\end{document}